# Active Disturbance Rejection Based Robust Trajectory Tracking Controller Design in State Space

Emre Sariyildiz[1], Rahim Mutlu[1], Chuanlin Zhang[2]
[1]School of Mechanical, Materials, Mechatronics and Biomedical Engineering
University of Wollongong, NSW, 2522, Australia
[2]College of Automation Engineering, Shanghai University of Electric Power, Shanghai, 200090, China
emre@uow.edu.au, rmutlu@uow.edu.au, clzhang@shiep.edu.cn

*Abstract-* **This paper proposes a new Active Disturbance Rejection (ADR) based robust trajectory tracking controller design method in state space. It can compensate not only matched but also mismatched disturbances. Robust state and control input references are generated in terms of a fictitious design variable, namely differentially flat output, and the estimations of disturbances by using Differential Flatness (DF) and Disturbance Observer (DOb). Two different robust controller design techniques are proposed by using Brunovsky canonical form and polynomial matrix form approaches. The robust position control problem of a two mass-spring-damper system is studied to verify the proposed ADR controllers.**

*Index Terms: Active Disturbance Rejection, Differential Flatness, Disturbance Observer, Matched and Mismatched Disturbances.*

## I. INTRODUCTION

Active Disturbance Rejection (ADR) control, in which robustness is achieved by directly cancelling disturbances, has several superiorities over Passive Disturbance Rejection (PDR) control, in which disturbances are suppressed via feedback regulation. For example, PDR controllers cannot react fast enough in the presence of a strong disturbance although it can be eventually suppressed [1]–[3]. Disturbance Observer (DOb) is an ADR control tool that is used to estimate disturbances and their successive time derivatives [3, 4]. If a system includes only matched disturbances, which act via the same channels as those of the control inputs, then the robustness can be simply achieved by feedbacking their estimations through control inputs [3] – [5]. However, many practical systems include mismatched disturbances, which act through different channels from those of the control inputs, as well as matched ones. Therefore, they cannot be suppressed, i.e., the robustness cannot be achieved, by using conventional ADR control [5] – [7]. Several studies have been carried out to suppress mismatched disturbances, e.g., sliding mode control, suppressing matched disturbances via ADR and mismatched disturbances via PDR control methods, and suppressing mismatched disturbances at output channel. However, they mainly suffer from design complexities and performance limitations [3, 5] – [8].

In this paper, a novel ADR-based robust trajectory tracking controller design method is proposed in state space. To achieve the performance goal, the state and control input references are generated in terms of the fictitious differentially flat output variable by using DF [9] – [11]. If DF is applicable (i.e., system dynamics is flat), then state and control input trajectories can be systematically generated in engineering applications such as under-actuated robots, compliant robots and unmanned aerial vehicles [12] – [14]. However, a conventional DF-based controller is sensitive to plant uncertainties and external disturbances; therefore, its stability and performance may significantly change in real implementations [11, 15]. To improve the robustness of a DF-based trajectory tracking controller, the state and control input references are systematically modified by using the estimations of disturbances and their successive time derivatives; i.e., not only the matched but also the mismatched disturbances are cancelled with their estimations in this paper. The robust trajectory tracking controllers are designed by using Brunovsky canonical form and polynomial matrix techniques. It is shown that the latter provides same performance and robustness with less computational load. By using the proposed robust controllers, a reference trajectory can be precisely tracked without requiring the exact dynamic models of the system and external disturbances. Therefore, they are applicable to many robust trajectory tracking control problems in different fields such as automotive and robotics. The proposal is verified by studying the robust trajectory tracking control problem of a two mass-spring-damper system.

The rest of the paper is organized as follows. In section II, $k^{th}$ order DOb is presented. In section III, two different design techniques are proposed for DF-based robust trajectory tracking control. In section IV, the proposed ADR controllers are applied to the trajectory tracking control problem of a two mass-spring-damper system. The paper ends with conclusion given in section V.

## II. $k^{th}$ ORDER DOb DESIGN IN STATE SPACE

Plant uncertainties and unmodeled / unknown disturbances can be estimated by using a zero-order, i.e., conventional, DOb [4, 16]. Not only disturbances but also their successive time derivatives can be similarly estimated by using a Higher Order DOb (HODOb) as follows:

Let us first describe the dynamic model of a system in state space by using

$$\dot{\mathbf{x}} = \mathbf{A}\mathbf{x} + \mathbf{B}u - \boldsymbol{\tau}_d \\ \dot{\mathbf{x}} = \mathbf{A}_n\mathbf{x} + \mathbf{B}_n u - \boldsymbol{\tau}_{dis} \quad (1)$$

where $\mathbf{A}$ and $\mathbf{A_n} \in \mathbb{R}^{p \times p}$ represent the exact and nominal system matrices, respectively; $\mathbf{B}$ and $\mathbf{B_n} \in \mathbb{R}^p$ represent the exact and nominal input vectors, respectively; $\mathbf{x}$ and $\dot{\mathbf{x}} \in \mathbb{R}^p$ represent the state vector of the system and its time derivative, respectively; $u \in \mathbb{R}$ represents the control input; $\boldsymbol{\tau}_d \in \mathbb{R}^p$ represents the unmodeled / unknown disturbance vector; and $\boldsymbol{\tau}_{dis} = \boldsymbol{\tau}_d + (\mathbf{A_n} - \mathbf{A})\mathbf{x} + (\mathbf{B_n} - \mathbf{B})u \in \mathbb{R}^p$ represents the vector of the disturbances due to plant uncertainties and $\boldsymbol{\tau}_d$.

It is noted that the parameters of the uncertain system, i.e., $\mathbf{A}$ and $\mathbf{B}$, may vary with time; yet the parameters of the nominal system, i.e., $\mathbf{A_n}$ and $\mathbf{B_n}$, are time invariant. Besides, the disturbance vectors, i.e., $\boldsymbol{\tau}_d$ and $\boldsymbol{\tau}_{dis}$, may include not only linear but also nonlinear disturbances.

Let us assume that the disturbance vector and its successive time derivatives are bounded such that

$$\left\| \overset{(j)}{\boldsymbol{\tau}_{dis}} \right\| \leq \delta_{\tau_{dis}}^j, \text{ where } j = 0, 1, 2, \cdots, k+1 \quad (2)$$

where $\|\bullet\|$ represents the norm of $\bullet$; $\overset{(j)}{\bullet}$ represents the $j^{th}$ derivative of $\bullet$; and $\delta_{\tau_{dis}}^j \geq 0 \in \mathbb{R}$ represents the upper bound of the $j^{th}$ derivative of the disturbance vector. The $k^{th}$ order DOb is designed in state space by using the following theorem.

***Theorem 1:*** The disturbance vector and its time derivatives up to order $k$ are estimated by using

$$\begin{aligned}
\hat{\boldsymbol{\tau}}_{dis} &= \hat{\mathbf{z}}_0 - L_0 \mathbf{x} \\
\hat{\dot{\boldsymbol{\tau}}}_{dis} &= \hat{\mathbf{z}}_1 - L_1 \mathbf{x} \\
&\vdots \\
\overset{(k-1)}{\hat{\boldsymbol{\tau}}}_{dis} &= \hat{\mathbf{z}}_{k-1} - L_{k-1} \mathbf{x} \\
\overset{(k)}{\hat{\boldsymbol{\tau}}}_{dis} &= \hat{\mathbf{z}}_k - L_k \mathbf{x}
\end{aligned} \quad (3)$$

where $\hat{\mathbf{z}}_j \in \mathbb{R}^p$ is the estimation of the $j^{th}$ auxiliary variable, i.e., $\mathbf{z}_j \in \mathbb{R}^p$; $\mathbf{x} \in \mathbb{R}^p$ is the state vector of the system which is given in Eq. (1); $\hat{\boldsymbol{\tau}}_{dis} \in \mathbb{R}^p$ is the estimation of the disturbance vector, i.e., $\boldsymbol{\tau}_{dis}$; $\hat{\dot{\boldsymbol{\tau}}}_{dis}, \hat{\ddot{\boldsymbol{\tau}}}_{dis}, \cdots, \overset{(k)}{\hat{\boldsymbol{\tau}}}_{dis} \in \mathbb{R}^p$ are the estimations of the disturbance vector's successive time derivatives, i.e., $\dot{\boldsymbol{\tau}}_{dis}, \ddot{\boldsymbol{\tau}}_{dis}, \cdots, \overset{(k)}{\boldsymbol{\tau}}_{dis} \in \mathbb{R}^p$, respectively; and $L_j \in \mathbb{R}$ represents the $j^{th}$ gain of DOb.

The estimations of the auxiliary variables are derived by integrating

$$\begin{aligned}
\dot{\hat{\mathbf{z}}}_0 &= -L_0 \hat{\mathbf{z}}_0 + \hat{\mathbf{z}}_1 + L_0 (\mathbf{A_n}\mathbf{x} + \mathbf{B_n}u + L_0\mathbf{x}) - L_1 \mathbf{x} \\
\dot{\hat{\mathbf{z}}}_1 &= -L_1 \hat{\mathbf{z}}_0 + \hat{\mathbf{z}}_2 + L_1 (\mathbf{A_n}\mathbf{x} + \mathbf{B_n}u + L_0\mathbf{x}) - L_2 \mathbf{x} \\
&\vdots \\
\dot{\hat{\mathbf{z}}}_{k-1} &= -L_{k-1} \hat{\mathbf{z}}_0 + \hat{\mathbf{z}}_k + L_{k-1} (\mathbf{A_n}\mathbf{x} + \mathbf{B_n}u + L_0\mathbf{x}) - L_k \mathbf{x} \\
\dot{\hat{\mathbf{z}}}_k &= -L_k \hat{\mathbf{z}}_0 + L_k (\mathbf{A_n}\mathbf{x} + \mathbf{B_n}u + L_0\mathbf{x})
\end{aligned} \quad (4)$$

where $\dot{\hat{\mathbf{z}}}_j \in \mathbb{R}^p$ represents the time derivative of the estimation of the $j^{th}$ auxiliary variable.

***Proof:*** Let us first design the auxiliary variables by using

$$\begin{aligned}
\mathbf{z}_0 &= L_0 \mathbf{x} + \boldsymbol{\tau}_{dis} \\
\mathbf{z}_1 &= L_1 \mathbf{x} + \dot{\boldsymbol{\tau}}_{dis} \\
&\vdots \\
\mathbf{z}_{k-1} &= L_{k-1} \mathbf{x} + \overset{(k-1)}{\boldsymbol{\tau}}_{dis} \\
\mathbf{z}_k &= L_k \mathbf{x} + \overset{(k)}{\boldsymbol{\tau}}_{dis}
\end{aligned} \quad (5)$$

Time derivative of Eq. (5) is derived as follows:

$$\begin{aligned}
\dot{\mathbf{z}}_0 &= -L_0 \mathbf{z}_0 + \mathbf{z}_1 + L_0 (\mathbf{A_n}\mathbf{x} + \mathbf{B_n}u + L_0\mathbf{x}) - L_1 \mathbf{x} \\
\dot{\mathbf{z}}_1 &= -L_1 \mathbf{z}_0 + \mathbf{z}_2 + L_1 (\mathbf{A_n}\mathbf{x} + \mathbf{B_n}u + L_0\mathbf{x}) - L_2 \mathbf{x} \\
&\vdots \\
\dot{\mathbf{z}}_{k-1} &= -L_{k-1}\mathbf{z}_0 + \mathbf{z}_k + L_{k-1}(\mathbf{A_n}\mathbf{x} + \mathbf{B_n}u + L_0\mathbf{x}) - L_k \mathbf{x} \\
\dot{\mathbf{z}}_k &= -L_k \mathbf{z}_0 + L_k (\mathbf{A_n}\mathbf{x} + \mathbf{B_n}u + L_0\mathbf{x}) + \overset{(k+1)}{\boldsymbol{\tau}}_{dis}
\end{aligned} \quad (6)$$

If Eq. (4) is subtracted from Eq. (6), then

$$\dot{\mathbf{e}}(t) = \boldsymbol{\Psi} \mathbf{e}(t) - \boldsymbol{\Gamma} \overset{(k)}{\boldsymbol{\tau}}_{dis} \quad (7)$$

where $\boldsymbol{\Psi} = \begin{bmatrix} -L_0 \mathbf{I_p} & \mathbf{I_p} & \mathbf{0_p} & \vdots & \mathbf{0_p} \\ -L_1 \mathbf{I_p} & \mathbf{0_p} & \mathbf{I_p} & \vdots & \mathbf{0_p} \\ \cdots & \cdots & \cdots & \vdots & \cdots \\ -L_k \mathbf{I_p} & \mathbf{0_p} & \mathbf{0_p} & \cdots & \mathbf{0_p} \end{bmatrix} \in \mathbb{R}^{(k+1)p \times (k+1)p}$; $\boldsymbol{\Gamma} = \begin{bmatrix} \mathbf{0_p} \\ \mathbf{0_p} \\ \vdots \\ \mathbf{I_p} \end{bmatrix} \in \mathbb{R}^{(k+1)p \times p}$;

$\mathbf{e}(t) = [\mathbf{z}_0 - \hat{\mathbf{z}}_0 \quad \mathbf{z}_1 - \hat{\mathbf{z}}_1 \quad \cdots \quad \mathbf{z}_k - \hat{\mathbf{z}}_k]^T$ represents the vector of the auxiliary variable estimation error; and $\mathbf{I_p}$ and $\mathbf{0_p} \in \mathbb{R}^{p \times p}$ represent identity and null matrices, respectively.

The dynamics of disturbance estimation is directly related to the eigenvalues of $\boldsymbol{\Psi}$, i.e., the bandwidth of DOb. The slowest eigenvalue of $\boldsymbol{\Psi}$ corresponds to the bandwidth of DOb and is derived by solving

$$\lambda(\boldsymbol{\Psi}) = \det(\lambda \mathbf{I}_{(k+1)p} - \boldsymbol{\Psi}) = (\lambda^{k+1} + L_k \lambda^k + \cdots + L_1 \lambda + L_0)^p = 0 \quad (8)$$

where $\lambda_{\min} = \lambda_0 \leq \lambda_1 \cdots \leq \lambda_k = \lambda_{\max}$ are the roots of Eq. (8).

If the gains of DOb are properly tuned so that the matrix $\boldsymbol{\Psi}$ is negative definite, then Eq. (7) satisfies the following inequality.

$$\|\mathbf{e}(t)\| \leq \exp(-\lambda_{\min} t) \|\mathbf{e}(t_0)\| + \lambda_{\min}^{-1} \delta_{\tau_{dis}}^{k+1} \quad (9)$$

Eq. (9) shows that any estimation error starts in a circular plane whose radius is $\|\mathbf{e}(t_0)\| + \lambda_{\min}^{-1} \delta_{\tau_{dis}}^{k+1}$ exponentially converges to a smaller circular plane whose radius is $\lambda_{\min}^{-1} \delta_{\tau_{dis}}^{k+1}$. The convergence rate and the accuracy of disturbance estimation are directly related to $\lambda_{\min}$, i.e., the bandwidth of DOb. One can simply improve the performance of disturbance estimation by increasing the bandwidth of DOb.

However, it is limited by practical constraints such as noise and sampling time in real implementations [17].

The performance of disturbance estimation is limited by the slowest eigenvalue of $\mathbf{\Psi}$. If the $k^{th}$ order DOb is designed by assigning $k+1$ repeated eigenvalues, i.e., $\lambda_0 = \lambda_1 = \cdots = \lambda_k = \lambda_{DOb}$, then Eq. (7) satisfies the following inequality.

$$\|\mathbf{e}(t)\| \leq k \exp(-\lambda_{DOb} t) \|\mathbf{e}(t_0)\| + k \lambda_{DOb}^{-1} \delta_{\tau_{dis}}^{k+1} \quad (10)$$

where $k = \|\exp(\mathbf{N}t)\|$ in which $\mathbf{N}$ represents Nilpotent matrix, i.e., $\exp(\mathbf{\Psi}t) = \exp(-\lambda_{DOb}t)\exp(\mathbf{N}t)$; and $\lambda_{DOb}$ is the bandwidth of DOb. Similarly, Eq. (10) shows that the convergence rate and the accuracy of disturbance estimation are improved as the bandwidth of DOb is increased. Q.E.D.

### III. DF-BASED ROBUST CONTROLLER DESIGN

If a system is flat, i.e., a linear system is controllable, then the state and control input references of its trajectory tracking controller can be systematically generated in terms of the fictitious differentially flat output variable and its successive time derivatives [9]. However, a conventional DF-based trajectory tracking controller requires the precise dynamic model of the system and is sensitive to external disturbances [11, 15]. Therefore, it is impractical in many applications such as robotics. In this section, DF-based robust trajectory tracking controllers are proposed by using Brunovsky canonical form and polynomial matrix techniques [10, 11].

*a) DF-based robust trajectory tracking controller design by using Brunovsky canonical form:*

If the nominal model of the system is flat, then Eq. (1) can be represented in Brunovsky canonical form by using [11]

$$\dot{\tilde{\mathbf{x}}} = \tilde{\mathbf{A}}_n \tilde{\mathbf{x}} + \tilde{\mathbf{B}}_n u - \tilde{\boldsymbol{\tau}}_{dis} \quad (11)$$

where $\tilde{\mathbf{A}}_n = \mathbf{T}\mathbf{A}_n\mathbf{T}^{-1} = \begin{bmatrix} \mathbf{0} & \mathbf{I} \\ \mathbf{a}_c^T & \end{bmatrix} \in \mathbb{R}^{p \times p}$ and $\tilde{\mathbf{B}}_n = \mathbf{T}\mathbf{B}_n = \begin{bmatrix} \mathbf{0} \\ 1 \end{bmatrix} \in \mathbb{R}^p$ represent the nominal system matrix and control input vector in Brunovsky canonical form, respectively; $\mathbf{T} \in \mathbb{R}^{p \times p}$ represent the transformation matrix; and $\tilde{\mathbf{x}} = \mathbf{T}\mathbf{x} = \begin{bmatrix} \tilde{x}_1 & \cdots & \tilde{x}_p \end{bmatrix}^T$ and $\tilde{\boldsymbol{\tau}}_{dis} = \mathbf{T}\boldsymbol{\tau}_{dis} = \begin{bmatrix} \tilde{d}_1 & \cdots & \tilde{d}_p \end{bmatrix}^T$ represent the state and disturbance vectors of the system in canonical form, respectively [18, 19].

The robust trajectory tracking controller can be designed by using the following theorem.

***Theorem 2:*** If the nominal model of the system which is given in Eq. (1) is controllable, then the robust control input can be designed by using

$$u_{r_1} = \overset{(p)}{y}_{DFO} + \mathbf{K}\left(\mathbf{x}^{ref} - \mathbf{x}\right) + \sum_{j=1}^{p} \widehat{\tilde{d}_j}^{(p-j)} - \mathbf{a}_c^T \mathbf{T} \mathbf{x}^{ref} \quad (12)$$

where $\mathbf{x}^{ref} = \mathbf{T}^{-1}\begin{bmatrix} y_{DFO} & \dot{y}_{DFO} + \hat{\tilde{d}}_1 & \cdots & \overset{(p-1)}{y}_{DFO} + \sum_{j=1}^{p-1}\widehat{\tilde{d}_j}^{(p-1-j)} \end{bmatrix}^T$ is the reference of the state vector; $y_{DFO}$ is the differentially flat output variable; $\overset{(k)}{\bullet}$ represents the $k^{th}$ derivative of $\bullet$; $\hat{\bullet}$ represents the estimation of $\bullet$; and $\mathbf{K}$ is the feedback control gain which is tuned by using the nominal model of the system in Eq. (1).

***Proof:*** Without any simplification, the state vector and control input of Eq. (11) can be derived by using

$$\tilde{\mathbf{x}} = \begin{bmatrix} \tilde{x}_1 \\ \tilde{x}_2 \\ \tilde{x}_3 \\ \vdots \\ \tilde{x}_p \end{bmatrix} = \begin{bmatrix} \tilde{x}_1 \\ \dot{\tilde{x}}_1 + \tilde{d}_1 \\ \dot{\tilde{x}}_2 + \tilde{d}_2 \\ \vdots \\ \dot{\tilde{x}}_{p-1} + \tilde{d}_{p-1} \end{bmatrix} = \begin{bmatrix} \tilde{x}_1 \\ \dot{\tilde{x}}_1 + \tilde{d}_1 \\ \ddot{\tilde{x}}_1 + \dot{\tilde{d}}_1 + \tilde{d}_2 \\ \vdots \\ \overset{(p-1)}{\tilde{x}}_1 + \sum_{j=1}^{p-1}\overset{(p-1-j)}{\tilde{d}}_j \end{bmatrix} \quad (13)$$

$$u = \dot{\tilde{x}}_p + \tilde{d}_p - \mathbf{a}_c^T \tilde{\mathbf{x}} = \overset{(p)}{\tilde{x}}_1 + \sum_{j=1}^{p} \overset{(p-j)}{\tilde{d}}_j - \mathbf{a}_c^T \tilde{\mathbf{x}} \quad (14)$$

The robust state and control input references are generated by applying $\tilde{x}_1 = y_{DFO}$ and the estimations of disturbances to Eq. (13) and Eq. (14). The robust trajectory tracking controller is designed by using a state feedback controller as shown in Eq. (12). Q.E.D.

A robust trajectory tracking controller can be systematically designed by using ***Theorem 2***. However, it is computationally demanding as the inverse of $\mathbf{T}$ is required in the design of the robust control input. A less computationally demanding robust trajectory tracking controller can be designed by using polynomial matrix approach.

*b) DF-based robust trajectory tracking controller design by using polynomial matrix form:*

Let us rewrite Eq. (1) in polynomial matrix form by using

$$\mathbf{A}_n(s)\mathbf{x}(s) = \mathbf{B}_n(s)u - \boldsymbol{\tau}_{dis}(s) \quad (15)$$

where $\mathbf{A}_n(s) \in \mathbb{R}^{p^* \times p^*}$ represents the polynomial system matrix; $\mathbf{B}_n(s) \in \mathbb{R}^{p^*}$ represents the polynomial control input vector; $\mathbf{x}(s) \in \mathbb{R}^{p^*}$ represents the state vector of the system; $\boldsymbol{\tau}_{dis}(s) \in \mathbb{R}^{p^*}$ represents the disturbance vector of the system; and $s$ represents differential operator. It is noted that the state space representation, which is given in Eq. (1), is a particular form of Eq. (15). Thus, the dimensions of Eq. (1) and Eq. (15), i.e., p and p*, can be different depending on the design.

The robust trajectory tracking controller can be designed by using the following theorem.

***Theorem 3:*** If the nominal model of the system which is given in Eq. (1) is controllable, then the robust control input can be designed by using

$$u_{r_2} = q_1(s)y_{DFO} + \mathbf{K}\left(\mathbf{x}^{ref} - \mathbf{x}\right) + \mathbf{q}_2^T(s)\hat{\boldsymbol{\tau}}_{dis}^m + \mathbf{q}_3^T(s)\hat{\boldsymbol{\tau}}_{dis}^{mm} \quad (16)$$

where $\mathbf{K}$ is the feedback control gain which is tuned by using the nominal model of the system in Eq. (1); $\hat{\boldsymbol{\tau}}_{dis}^m$ and $\hat{\boldsymbol{\tau}}_{dis}^{mm}$ represent the estimations of the matched and mismatched disturbances, respectively; and

$$\mathbf{x}^{ref}(s) = \tilde{\mathbf{p}}(s) y_{DFO} = \mathbf{p}_1(s) y_{DFO} + \mathbf{P}_2(s) \hat{\boldsymbol{\tau}}_{dis}^{mm}(s) \quad (17)$$

$\tilde{\mathbf{p}}(s) y_{DFO}$ is derived by solving

$$\mathbf{c}^T(s) \mathbf{A_n}(s) \tilde{\mathbf{p}}(s) y_{DFO} + \mathbf{c}^T(s) \hat{\boldsymbol{\tau}}_{dis}^{mm}(s) = 0 \quad (18)$$

where $\mathbf{c}(s) \in \mathbb{R}^{p^*}$ is orthogonal to $\mathbf{B_n}(s)$, i.e., $\mathbf{c}^T(s)\mathbf{B_n}(s) = 0$.

$q_1(s)$, $\mathbf{q_2}(s)$ and $\mathbf{q_3}(s)$ are obtained by using

$$\begin{aligned} q_1(s) &= \left(\mathbf{B_n}^T(s)\mathbf{B_n}(s)\right)^{-1} \mathbf{B_n}^T(s) \mathbf{A_n}(s) \mathbf{p}_1(s) \\ \mathbf{q_2}(s) &= \left(\mathbf{B_n}^T(s)\mathbf{B_n}(s)\right)^{-1} \mathbf{B_n} \\ \mathbf{q_3}(s) &= \left(\mathbf{B_n}^T(s)\mathbf{B_n}(s)\right)^{-1} \mathbf{P}_2^T(s) \mathbf{A_n}^T(s) \mathbf{B_n}(s) \end{aligned} \quad (19)$$

**Proof:** Since the nominal model of the system is controllable, its states and control input can be defined in terms of differentially flat output variable. Eq. (15) can be rewritten as follows:

$$\mathbf{A_n}(s)\tilde{\mathbf{p}}(s) y_{DFO} + \mathbf{d}(s) y_{DFO} = \mathbf{B_n}(s) \tilde{q}(s) y_{DFO} \quad (20)$$

where $\mathbf{x}(s) = \tilde{\mathbf{p}}(s) y_{DFO}$, $u = \tilde{q}(s) y_{DFO}$, and $\boldsymbol{\tau}_{dis}(s) = \mathbf{d}(s) y_{DFO}$ in which $\mathbf{d}(s) = \mathbf{B_n}(s)\tilde{q}(s) - \mathbf{A_n}(s)\tilde{\mathbf{p}}(s) \in \mathbb{R}^{p^*}$.

Let us separate the matched and mismatched disturbances of $\boldsymbol{\tau}_{dis}(s)$ and $\mathbf{d}(s)$ by using

$$\begin{aligned} \boldsymbol{\tau}_{dis}(s) &= \boldsymbol{\tau}_{dis}^m(s) + \boldsymbol{\tau}_{dis}^{mm}(s) \\ \mathbf{d}(s) &= \mathbf{d}^m(s) + \mathbf{d}^{mm}(s) \end{aligned} \quad (21)$$

If Eq. (20) is multiplied by $\mathbf{c}^T(s)$, which is orthogonal to $\mathbf{B_n}(s)$, from the left side and Eq. (21) is substituted into the disturbance vector, then Eq. (20) is derived as follows:

$$\mathbf{c}^T(s)\mathbf{A_n}(s)\mathbf{w}(s) = 0 \quad (22)$$

where $\mathbf{w}(s) = \tilde{\mathbf{p}}(s) + \mathbf{A_n}^{-1}(s)\mathbf{d}^{mm}(s) \in \mathbb{R}^{p^*}$.

Eq. (22) shows that the polynomial $\mathbf{w}(s)$ is orthogonal to $\mathbf{A_n}^T(s)\mathbf{c}(s)$. It can be obtained by using

$$\mathbf{w}(s) = \mathbf{R}\left(\mathbf{r}, \frac{\pi}{2}\right) \mathbf{A_n}^T(s) \mathbf{c}(s) \quad (23)$$

where $\mathbf{R}(\mathbf{r},\theta) \in \mathbb{R}^{p^* \times p^*}$ represents an orthogonal rotational matrix in which $\mathbf{r} \in \mathbb{R}^{p^*}$ represents the axis of rotation and $\theta \in \mathbb{R}$ represents the angle of rotation. Eq. (23) shows that the polynomial $\mathbf{w}(s)$ has no unique solution.

Since $\mathbf{A_n}(s)$ is full rank, the polynomial $\tilde{\mathbf{p}}(s)$ can be obtained by using

$$\tilde{\mathbf{p}}(s) = \mathbf{w}(s) - \mathbf{A_n}^{-1}(s)\mathbf{d}^{mm}(s) \quad (24)$$

Eq. (19) can be directly derived by multiplying Eq. (20) with $\left(\mathbf{B_n}^T(s)\mathbf{B_n}(s)\right)^{-1} \mathbf{B_n}^T(s)$ from the left side.

Hence, states and control input are derived in terms of the disturbance vector and differentially flat output variable. The robust state and control input references can be generated by using the estimations of disturbances via DOb.  Q.E.D.

*c) Stability Analysis:*

The fundamental idea behind **Theorem 2** and **Theorem 3** is that if the state vector of the system is properly modified by using the mismatched disturbances, then a system model which suffers from only matched disturbances is achieved. The robust trajectory tracking controllers are designed by generating the references of the reconstructed state vectors, suppressing the matched disturbances by feedbacking their estimations and designing a state feedback controller.

For example, without any simplification, Eq. (11) can be rewritten by using

$$\dot{\tilde{\boldsymbol{\xi}}} = \tilde{\mathbf{A}}_n \tilde{\boldsymbol{\xi}} + \tilde{\mathbf{B}}_n \left( u - d_{\tilde{\xi}} \right) \quad (25)$$

where $\tilde{\boldsymbol{\xi}} = \begin{bmatrix} \tilde{\xi}_1 & \tilde{\xi}_2 & \tilde{\xi}_3 & \cdots & \tilde{\xi}_p \end{bmatrix}^T$ in which $\tilde{\xi}_1 = \tilde{x}_1$, $\tilde{\xi}_2 = \tilde{x}_2 - \tilde{d}_1$, $\tilde{\xi}_3 = \tilde{x}_3 - \dot{\tilde{d}}_1 - \tilde{d}_2, \cdots$, and $\tilde{\xi}_p = \tilde{x}_p - \sum_{j=1}^{p-1} \tilde{d}_j^{(p-1-j)}$ ; and $d_{\tilde{\xi}} = \mathbf{a}_c^T \left(\tilde{\boldsymbol{\xi}} - \tilde{\mathbf{x}}\right) + \sum_{j=1}^{p} \tilde{d}_j^{(p-j)}$ represents the matched disturbance.

More generally, Eq. (15) can be rewritten by reconstructing the state vector of the system as follows:

$$\mathbf{A}_n(s) \boldsymbol{\xi}(s) = \mathbf{B}_n(s) u - \boldsymbol{\tau}_{dis}^m \quad (26)$$

where $\boldsymbol{\xi}(s) = \mathbf{x}(s) + \mathbf{A}_n^{-1}(s) \boldsymbol{\tau}_{dis}^{mm}$; and $\boldsymbol{\tau}_{dis}^m$ and $\boldsymbol{\tau}_{dis}^{mm}$ represent the matched and mismatched disturbance vectors, respectively. Similarly, Eq. (26) can be represented in state space by using

$$\dot{\boldsymbol{\xi}} = \mathbf{A}_n \boldsymbol{\xi} + \mathbf{B}_n u - \boldsymbol{\tau}_{dis}^m \quad (27)$$

In **Theorem 2** and **Theorem 3**, the state and control input references are generated by using the reconstructed state space representations which are given in Eq. (25) and Eq. (27). The following theorem proves the stability of the proposed robust controllers.

**Theorem 4:** If the robust trajectory tracking controllers are designed by using **Theorem 2** or **Theorem 3**, then all states of the system are uniformly ultimately bounded with respect to the set

$$\begin{aligned} \Omega_1 &= \left\{ \boldsymbol{\xi}(t) \in \mathbb{R}^p : \|\boldsymbol{\xi}(t)\|^2 \leq \frac{\varphi}{\ell} \right\} \\ \Omega/\Omega_1 &= \left\{ \boldsymbol{\xi}(t) \in \mathbb{R}^p : \|\boldsymbol{\xi}(t)\|^2 > \frac{\varphi}{\ell} \right\} \end{aligned} \quad (28)$$

where $\ell < \lambda_{\min}(\mathbf{Q}) - 1$; $\varphi = \left(|\lambda(\mathbf{P})|_{\max}\right)^2 \left(\delta_{\boldsymbol{\tau}_{dis}^m}^m\right)^2$ in which $\delta_{\boldsymbol{\tau}_{dis}^m}^m$ represents the upper bound of the matched disturbance estimation error and $|\lambda(\mathbf{P})|_{\max}$ represents the maximum norm of the eigenvalues of $\mathbf{P}$; and $\mathbf{Q}$ and $\mathbf{P}$ are positive definite matrices which satisfy

$$\left(\mathbf{A_n}-\mathbf{B_n K}\right)^T \mathbf{P} + \mathbf{P}\left(\mathbf{A_n}-\mathbf{B_n K}\right) = -\mathbf{Q} \qquad (29)$$

**Proof:** Let us design the Lyapunov function candidate by using

$$V = \xi^T \mathbf{P} \xi \qquad (30)$$

When $\mathbf{B_n}u = -\mathbf{B_n K}\xi + \hat{\boldsymbol{\tau}}_{\mathbf{dis}}^{\mathbf{m}}$, the derivate of Eq. (30) is derived as follows:

$$\dot{V} = -\xi^T \mathbf{Q}\xi + 2\xi^T \mathbf{P}\left(\hat{\boldsymbol{\tau}}_{\mathbf{dis}}^{\mathbf{m}} - \boldsymbol{\tau}_{\mathbf{dis}}^{\mathbf{m}}\right) \qquad (31)$$

where $\hat{\boldsymbol{\tau}}_{\mathbf{dis}}^{\mathbf{m}}$ represents the estimation of the matched disturbance vector, i.e., $\boldsymbol{\tau}_{\mathbf{dis}}^{\mathbf{m}}$.

Eq. (31) satisfies

$$\begin{aligned}\dot{V} &\leq -\lambda_{\min}(\mathbf{Q})\|\xi\|^2 + 2\|\xi\|\left|\lambda(\mathbf{P})\right|_{\max}\left|\delta_{\boldsymbol{\tau}_{\mathbf{dis}}^{\mathbf{m}}}\right| \\ &\leq -\lambda_{\min}(\mathbf{Q})\|\xi\|^2 + \|\xi\|^2 + \left(\left|\lambda(\mathbf{P})\right|_{\max}\left|\delta_{\boldsymbol{\tau}_{\mathbf{dis}}^{\mathbf{m}}}\right|\right)^2 \\ &\leq -\left(\lambda_{\min}(\mathbf{Q})-1\right)\|\xi\|^2 + \varphi \end{aligned} \qquad (32)$$

Eq. (28) and Eq. (32) show that the time derivative of the Lyapunov function is negative outside of the compact set $\Omega_1$. Therefore, any states start in $\Omega/\Omega_1$ ultimately enter in $\Omega_1$.

Q.E.D.

## IV. ROBUST POSITION CONTROL OF A TWO MASS-SPRING-DAMPER SYSTEM

In this section, robust trajectory tracking controllers are designed for a two mass-spring-damper system, which is illustrated in Fig. 1, by using **Theorem 2** and **Theorem 3**. In this figure, $m_i$ represents the $i^{th}$ mass; $b_i$ represents the $i^{th}$ viscous friction coefficient; $k$ represents the stiffness of the spring; $q_i$, $\dot{q}_i$, and $\ddot{q}_i$ represent the position, velocity and acceleration of the $i^{th}$ mass, respectively; and $F_{in}$ and $F_{ext}$ represent input and output forces, such as motor torque and external load, respectively.

The dynamic equations of the two mass-spring-damper system can be directly derived from Fig.1 as follows:

$$\begin{aligned} m_{1n}\ddot{q}_1 + b_{1n}\dot{q}_1 &= F_{in} - k_n(q_1-q_2) - d_1 \\ m_{2n}\ddot{q}_2 + b_{2n}\dot{q}_2 &= k_n(q_1-q_2) - d_2 \end{aligned} \qquad (33)$$

where $m_{\bullet n}$, $b_{\bullet n}$ and $k_n$ represent the nominal parameters of $m_\bullet$, $b_\bullet$ and $k$, respectively; and $d_1$ and $d_2$ represent the matched and mismatched disturbances, i.e.,

$$\begin{aligned} d_1 &= (m_1-m_{1n})\ddot{q}_1 + (k-k_n)(q_1-q_2) + f_{ud1} \\ d_2 &= F_{ext} + (m_2-m_{2n})\ddot{q}_2 + (k_n-k)(q_1-q_2) + f_{ud2} \end{aligned} \qquad (34)$$

where $f_{ud*}$ represents any linear and nonlinear unmodeled / unknown disturbances.

Without any simplification, the dynamic model of the system can be represented in state space as follows:

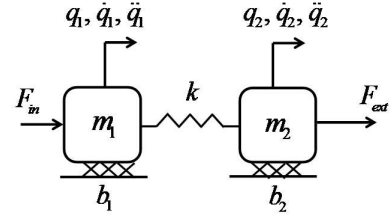

Fig.1: Two mass-spring-damper system.

$$\dot{\mathbf{x}} = \mathbf{A_n}\mathbf{x} + \mathbf{B_n}u - \boldsymbol{\tau}_{\mathbf{dis}} \qquad (35)$$

where $\mathbf{A_n} = \begin{bmatrix} 0 & 1 & 0 & 0 \\ -\frac{k_n}{m_{1n}} & -\frac{b_{1n}}{m_{1n}} & \frac{k_n}{m_{1n}} & 0 \\ 0 & 0 & 0 & 1 \\ \frac{k_n}{m_{2n}} & 0 & -\frac{k_n}{m_{2n}} & -\frac{b_{2n}}{m_{2n}} \end{bmatrix}$, $\mathbf{B_n} = \begin{bmatrix} 0 \\ \frac{1}{m_{1n}} \\ 0 \\ 0 \end{bmatrix}$, $\boldsymbol{\tau}_{\mathbf{dis}} = \begin{bmatrix} 0 \\ \frac{d_1}{m_{1n}} \\ 0 \\ \frac{d_2}{m_{2n}} \end{bmatrix}$,

$\mathbf{x} = \begin{bmatrix} q_1 & \dot{q}_1 & q_2 & \dot{q}_2 \end{bmatrix}^T, u = F_{in}$.

Eq. (35) shows that the system suffers from matched and mismatched disturbances in the second and fourth channels, respectively. Since the nominal model of the system is controllable, i.e., $\boldsymbol{\Gamma} = \begin{bmatrix} \mathbf{B_n} & \mathbf{A_n B_n} & \mathbf{A_n^2 B_n} & \mathbf{A_n^3 B_n} \end{bmatrix}$ is full rank, the robust trajectory tracking controller can be designed by using either **Theorem 2** or **Theorem 3**. The former can be systematically applied by deriving the Brunovsky canonical form of Eq. (35).

Let us focus on designing the robust trajectory tracking controller by using **Theorem 3**. Eq. (33) can be represented in polynomial matrix form by using

$$\mathbf{A_n}(s)\mathbf{x}(s) + \boldsymbol{\tau}_{\mathbf{dis}}(s) = \mathbf{B_n}(s)u \qquad (36)$$

where $\mathbf{A_n}(s) = \begin{bmatrix} m_{1n}s^2 + b_{1n}s + k_n & -k_n \\ -k_n & m_{2n}s^2 + b_{2n}s + k_n \end{bmatrix}$, $\mathbf{B_n}(s) = \begin{bmatrix} 1 \\ 0 \end{bmatrix}$, $u = F_{in}$,

$\mathbf{x}(s) = \begin{bmatrix} q_1 & q_2 \end{bmatrix}^T$, $\boldsymbol{\tau}_{\mathbf{dis}}(s) = \begin{bmatrix} d_1 & d_2 \end{bmatrix}^T$ and $s$ represents differential operator.

If Eq. (36) is multiplied with $\mathbf{c}^T(s) = \begin{bmatrix} 0 & 1 \end{bmatrix}$, which is orthogonal to $\mathbf{B_n}(s)$, from the left side, then we derive

$$\mathbf{c}^T(s)\mathbf{A_n}(s)\mathbf{x}(s) + \mathbf{c}^T(s)\boldsymbol{\tau}_{\mathbf{dis}}(s) = 0 \qquad (37)$$

where $\mathbf{c}^T(s)\boldsymbol{\tau}_{\mathbf{dis}}(s) = d_2$ is the mismatched disturbance of the system and $\mathbf{c}^T(s)\mathbf{A_n}(s) = \begin{bmatrix} -k_n & m_{2n}s^2 + b_{2n}s + k_n \end{bmatrix}$.

If $\mathbf{x}(s) = \tilde{\mathbf{p}}(s)y_{DFO} = \begin{bmatrix} p_1(s) & p_2(s) \end{bmatrix}^T y_{DFO}$ is substituted into Eq. (37), then we obtain

$$\begin{bmatrix} -k_n & m_{2n}s^2 + b_{2n}s + k_n \end{bmatrix}\begin{bmatrix} p_1(s) \\ p_2(s) \end{bmatrix} y_{DFO} + d_2 = 0 \qquad (38)$$

where $y_{DFO}$ represents the differentially flat output variable.

ASME Journal of Journal of Dynamic Systems, Measurement, and Control

TABLE I: Parameters of Simulation.

| Parameters | Description | Values |
|---|---|---|
| $m_1$ | Mass of motor | 0.1 kg |
| $m_2$ | Mass of link | 0.25 kg |
| $b_1$ and $b_2$ | Viscous friction coef. | 2.5 Ns/m |
| $b_{12}$ | Viscous friction coef. | 1.25 Ns/m |
| $k$ | Spring Stiffness | 100 N/m |

As shown in *Theorem 3*, there is no unique solution for $\tilde{\mathbf{p}}(s)$, i.e., Eq. (38). If it is assumed that $p_2(s) = k_n$, then the state vector of the system is derived as follows:

$$\mathbf{x}(s) = \tilde{\mathbf{p}}(s) y_{DFO} = \mathbf{p_1}(s) y_{DFO} + \mathbf{P_2}(s) \boldsymbol{\tau}_{dis}^{mm}(s) \quad (39)$$

where $\mathbf{p_1}(s) = \begin{bmatrix} m_{2n}s^2 + b_{2n}s + k_n \\ k_n \end{bmatrix}$, $\mathbf{P_2}(s) = \begin{bmatrix} 0 & 1 \\ 0 & k_n \\ 0 & 0 \end{bmatrix}$ and $\boldsymbol{\tau}_{dis}^{mm}(s) = \begin{bmatrix} 0 \\ d_2 \end{bmatrix}$.

The control input can be directly derived by using *Theorem 3* and Eq. (39) as follows:

$$u = q_1(s) y_{DFO} + \mathbf{q_2}^T(s) \boldsymbol{\tau}_{dis}^{m} + \mathbf{q_3}^T(s) \boldsymbol{\tau}_{dis}^{mm} \quad (40)$$

where $q_1(s) = m_{1n}m_{2n}s^4 + (m_{1n}b_{2n} + m_{2n}b_{1n})s^3 + (b_{1n}b_{2n} + k_n(m_{1n} + m_{2n}))s^2 + k_n(b_{1n} + b_{2n})s$; $\mathbf{q_2}^T(s) = \begin{bmatrix} 1 & 0 \end{bmatrix}$; $\mathbf{q_3}^T(s) = \begin{bmatrix} 0 & \frac{m_{1n}}{k_n}s^2 + \frac{b_{1n}}{k_n}s + 1 \end{bmatrix}$; $\boldsymbol{\tau}_{dis}^{m}(s) = \begin{bmatrix} d_1 & 0 \end{bmatrix}^T$; and $\boldsymbol{\tau}_{dis}^{mm}(s) = \begin{bmatrix} 0 & d_2 \end{bmatrix}^T$.

The state and control input references can be generated by applying the estimations of disturbances to Eq. (39) and Eq. (40). The differentially flat output variable is designed in terms of control goal. For example, to follow the trajectory of the second mass, the differentially flat output variable is designed as follows:

$$y_{DFO} = q_2^{des} / k_n \quad (41)$$

where $q_2^{des}$ represents the desired $q_2$.

Let us now validate the proposed robust controllers by giving the simulation results of the position control of a two mass-spring-damper system. In simulations, the position of the second mass (e.g. link of a compliant actuator) is controlled when step and sinusoidal reference inputs are applied. It is assumed that the plant parameters are uncertain, i.e., $m_{1n} = 0.65 m_1, m_{2n} = 0.35 m_2, b_{1n} = b_{2n} = 0$ and $k_n = 2.65 k$; and an external disturbance is applied between 2.5 and 10 seconds by using $f_{ext} = 25(\sin(2\pi t)\cos(6\pi t))$. The parameters of the simulation are given in Table I.

DF-based trajectory tracking controllers are designed by using the following steps:

- First, all disturbances are neglected. In regulation control, the state feedback controller is designed as $\mathbf{K} = \begin{bmatrix} -167.7321 & 7.15 & 179.8047 & -5.3794 \end{bmatrix}$ so that the double poles of the nominal system are placed at -25 and -30; however, in trajectory tracking control, the state feedback controller is designed as $\mathbf{K} = \begin{bmatrix} 714.6429 & 14.3 & -521.4825 & -0.1349 \end{bmatrix}$ so that the double poles of the nominal system are placed at -50 and -60.
- The state and control input references are generated in terms of differentially flat output variable. Hence, the conventional DF-based trajectory tracking controller is designed in state space.
- The matched and mismatched disturbances and their first and second order derivatives are estimated by using the second order DOb. The gains of DOb are tuned by placing all eigenvalues of $\boldsymbol{\Psi}$ at 1000, i.e., setting the bandwidth of DOb at 1000 rad/s.
- Robust state and control input references are generated by using the estimations of disturbances as shown in *Theorem 2* and *Theorem 3*.

Regulation and trajectory tracking control results are illustrated in Fig. 2 when DF-based position controllers are implemented. Fig. 2a and Fig. 2b show that the conventional DF-based position controller is sensitive to parametric uncertainties and external disturbances. The second mass can precisely track the step and sinusoidal references when $\boldsymbol{\tau}_{dis} = \mathbf{0}$. However, not only the performance but also the stability of the conventional DF-based position controller may significantly deteriorate by disturbances. The regulation and trajectory tracking control results are respectively illustrated in Fig. 2c and Fig. 2d when the proposed DF-based robust position controllers are implemented. It is clear from these figures that the DF-based robust position controllers can suppress parametric uncertainties and external disturbances. The step and sinusoidal references can be tracked without requiring the precise dynamic models of the system and external disturbances.

Disturbances and their estimations are illustrated in Fig. 3 when the proposed DF-based robust position controller is implemented. First, it is assumed that the dynamic model of the system is precisely known; i.e., the system suffers from only the mismatched external disturbances. Fig. 3a shows that DOb can work as a force/torque sensor and estimate external load when the dynamic model of the plant is precisely known. However, if the system suffers from not only external disturbances but also plant uncertainties, then the dynamic model includes matched disturbances as well as mismatched ones. The matched and mismatched disturbances and their estimations are illustrated in Fig. 3b. The first and second order derivatives of the disturbances and their estimations are illustrated in Fig. 3c and Fig. 3d, respectively. It is clear from these figures that the proposed second order DOb can precisely estimate disturbances and their first and second order derivatives.

To minimize the influence of disturbance estimation, the dynamics of DOb should be tuned faster than that of the performance controller, i.e., state feedback controller. However, the bandwidth of disturbance estimation is limited by practical constraints such as noise and sampling time. In other words, there is a trade-off between the robustness and noise sensitivity of the proposed DF-based robust trajectory tracking controller.

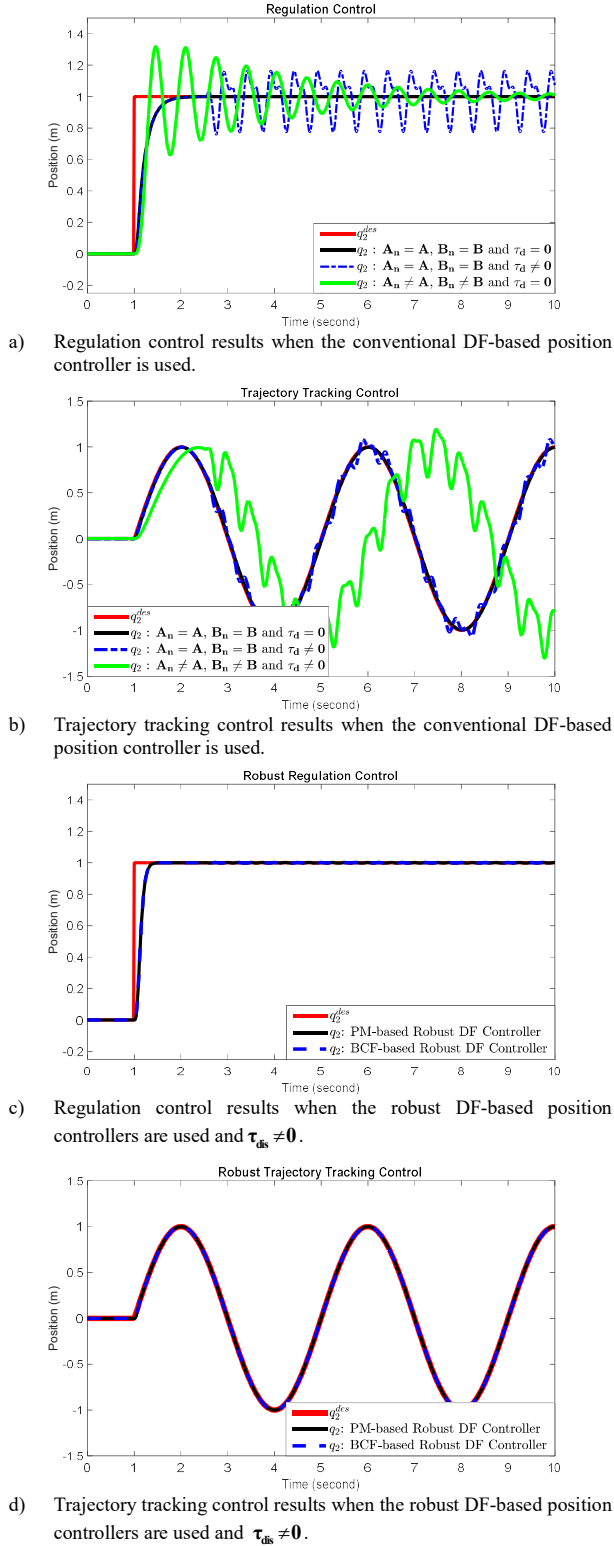

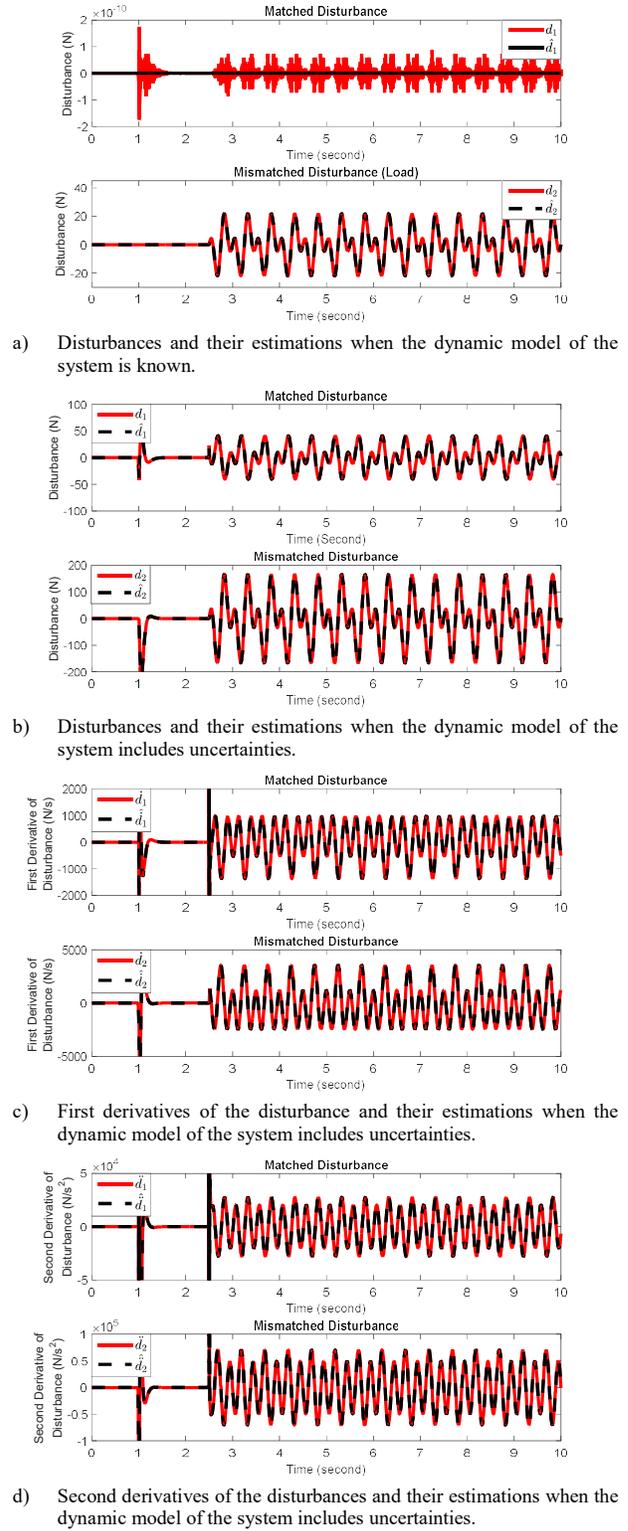

a) Regulation control results when the conventional DF-based position controller is used.

b) Trajectory tracking control results when the conventional DF-based position controller is used.

c) Regulation control results when the robust DF-based position controllers are used and $\tau_{dis} \neq 0$.

d) Trajectory tracking control results when the robust DF-based position controllers are used and $\tau_{dis} \neq 0$.

Fig. 2: Regulation and trajectory tracking control results. PM-based and BCF-based robust DF controllers represent the Polynomial Matrix-based and Brunovsky Canonical Form-based robust differential flatness controllers, respectively.

a) Disturbances and their estimations when the dynamic model of the system is known.

b) Disturbances and their estimations when the dynamic model of the system includes uncertainties.

c) First derivatives of the disturbance and their estimations when the dynamic model of the system includes uncertainties.

d) Second derivatives of the disturbances and their estimations when the dynamic model of the system includes uncertainties.

Fig. 3: Matched and mismatched disturbances and their estimations.

Simulation results show that same position control performances can be obtained when the Brunovsky canonical form and polynomial matrix based robust trajectory tracking controllers are implemented. However, the former has more computational load than the latter.

## V. CONCLUSION

In this paper, a new ADR-based robust trajectory tracking controller design method, which suppresses not only matched but also mismatched disturbances, is proposed by using DF and DOb in state space. The robust state and control input references are systematically generated in terms of differentially flat output variable, estimations of disturbances and their successive time derivatives by using Brunovsky canonical form and polynomial matrix form based design techniques. They provide same performance with different computational loads. By using the proposed robust controllers, reference trajectories can be precisely tracked when systems suffer from plant uncertainties and external disturbances. The validity of the proposal is verified by giving simulation results of a two mass-spring-damper system.